\documentclass[11pt,a4paper]{article}
\usepackage{amsmath}
\usepackage{amssymb}
\usepackage[latin1]{inputenc}
\usepackage[tmargin=3cm,lmargin=3cm,rmargin=3cm,bmargin=3cm]{geometry}
\usepackage{fancyhdr}
\usepackage{authblk}	

\title{\textbf{The Effects of Majorana Phases in Estimating the Masses of Neutrinos}}
\author{Ng. K. Francis \thanks{ngkf2010@gmail.com}}
\affil{Department of Physics, Tezpur University, Tezpur-784028, India}
\date{}

\begin{document}

	\Large 
	\maketitle
	\rhead{\thepage} 	
	\begin{abstract}
		 Majorana CP violating phases coming from heavy right-handed Majorana mass matrices ($M_{RR}$) are considered to estimate the masses of neutrinos.The effects of phases on quasi-degenerate neutrinos mass matrix obeying $\mu$-$\tau$ symmetry predicts the results consistent with observations for (i) solar mixing angle($\theta_{12}$) below TBM, (ii) absolute neutrino mass parameters[$m_{ee}$] in neutrinoless double beta ($0\nu\beta\beta$) decay, and (iii) cosmological upper bound $\sum_{i}m_{i}$. Analysis is carried out through parameterization of light left-handed Majorana neutrino matrices $(m_{LL})$ using only two unknown parameters $(\epsilon,\eta)$ within $\mu$-$\tau$ symmetry. We consider the charge lepton and up quark matrices as diagonal form of Dirac neutrino mass matrix $(m_{LR})$, and $m_{RR}$ are genrated using $m_{LL}$ through inversion of Type-I seesaw formula. The analysis shows that the masses of neutrinos are in agreement with the upper bound from cosmology and neutrinoless double beta decay. The results presented in this article will have important implications in discriminating the neutrinos mass models. \\
		\begin{description}
			\item[Keywords]
			majorana phases, absolute neutrino masses, QDN models. 
			\item[PACS Numbers]
			14.60. Pq; 12.15. Ff; 13.40. Em.
		\end{description}
	\end{abstract}
	\newpage
	\maketitle
	
	\section{\label{sec:level1}\Large{Introduction}}
	Since the present neutrino oscillation data[1] on neutrino mass parameters are not sufficient to predict the three absolute neutrino masses in the case of quasi-degenerate neutrino (QDN) mass models[2-8], such mass scale is usually taken as input ranging from 0.1 to 0.4 eV in most of the theoretical calculations[9]. As the latest cosmological tightest upper bound on the sum of the three absolute masses is $\sum m_{i}\leq0.28$ eV[11], larger value of neutrino mass $m_{3}\geq0.1$ eV in QDN models, has been disfavoured. The upper bound on $m_{ee}\geq0.2$ eV in neutrinoless double beta decay ($0\nu\beta\beta$) [11] also disfavour larger values of neutrino mass eigenvalues with same CP-parity. Some important points for further investigations in QDN models for both NH and IH patterns are searches for QDN models which can accomodate lower values of absolute neutrino masses $m_{3}\geq0.09$ eV, solar mixing angle which is lower than tribimaximal mixing (TBM)[12] and effects of CP-phases on neutrino masses. In this paper, we introduce a general classification for QDN models based on their CP-parity patterns and then parameterize the mass matrix within $\mu$-$\tau$ symmetry, and finally numerical calculations are carried out.
	\section{\label{sec:level2}\Large{Parameterizations of neutrino mass matrix}}
	A general $\mu$-$\tau$ symmetric neutrino mass matrix[13,14] with its four unknown independent matrix elements, requires at least four independent equations for realistic numerical solution.
	\begin{equation}
	m_{LL}=
	\begin{bmatrix}
	m_{11} & m_{12} & m_{13}\\
	m_{21} & m_{22} & m_{23}\\
	m_{31} & m_{32} & m_{33}
	\end{bmatrix}
	\end{equation}
	The three mass eigenvalues, $m_{i}$ and solar mixing angles, $\theta_{12}$ are given by:
	\begin{equation*}
	m_{1}=m_{11}-\sqrt{2}\tan\theta_{12}m_{12} \mbox{,} \quad m_{2}=m_{11}+\sqrt{2}\cot\theta_{12}m_{12} \mbox{,} \quad  m_{3}=m_{22}-m_{23}
	\end{equation*}
	\begin{equation}
	\tan 2\theta_{12}=\frac{2\sqrt{2}m_{12}}{m_{11}-m_{22}-m_{23}}
	\end{equation} 
	The observed mass-squared differences are calculated as
	$\triangle m_{12}^{2}=m_{2}^{2}-m_{1}^{2}>0$, $\triangle m_{32}^{2}=\left|m_{3}^{2}-m_{2}^{2}\right|$. In the basis where charged lepton mass matrix is diagonal, we have the leptonic mixing matrix, $U_{PMNS}=U$, where
	\begin{equation}
	U_{PMNS}=
	\begin{bmatrix}
	\cos\theta_{12} & \sin\theta_{12} & 0\\
	\frac{\sin\theta_{12}}{\sqrt{2}} & \frac{\cos\theta_{12}}{\sqrt{2}} & -\frac{1}{\sqrt{2}}\\
	\frac{\sin\theta_{12}}{\sqrt{2}} & \frac{\cos\theta_{12}}{\sqrt{2}} & \frac{1}{2}
	\end{bmatrix}
	\end{equation}
	The mass parameters $m_{ee}$ in $\nu\beta\beta$ decay and the sum of the absolute neutrino masses in WMAP cosmological bound $\sum m_{i}$, are given respectively by: 
	\begin{align*}
	m_{ee}=\left|m_{1}U_{e1}^{2}+m_{2}U_{e2}^{2}+m_{3}U_{e3}^{2}\right|
	\end{align*}
	\begin{align*}
	\mbox{and }m_{cosmos}=m_{1}+m_{2}+m_{3}
	\end{align*}
	
	A general classification for three-fold quasi-degenerate neutrino mass models[13] with respect to Majorana CP-phases in their three mass eigenvalues, is adopted here. Diagonalisation of left-handed Majorana neutrino mass matrix $m_{LL}$ in equation (1) is given by $m_{LL}=UDU^{T}$, where U is the diagonalising matrix in equation (4) and Diag$=D(m_{1},m_{2}e^{i\alpha},m_{3}e^{i\beta})$ is the diagonal matrix with two unknown Majorana phases $(\alpha,\beta)$. In the basis where charged lepton mass matrix is diagonal, the leptonic mixing matrix is given by $U=U_{PMNS}$[14]. We, then, adopt the following classificaion according to their CP-parity patterns in the mass eigenvalues $m_{i}$ namely \textbf{Type IA: (+-+)} for D$=Diag(m_{1},-m_{2},m_{3})$; \textbf{Type IB: (+++)} for D$=Diag(m_{1},m_{2},m_{3})$ and \textbf{Type IC: (++-)} for D$=Diag(m_{1},-m_{2},-m_{3})$ respectively. We now introduce the following parameterization for $\mu$-$\tau$ symmetric neutrino mass matrices $\mu_{LL}$ which could satisfy the above classifications[13].\\
	\begin{table}[b]
		\begin{center}
			\begin{tabular}{cccccc}\hline\hline
				\multicolumn{1}{c}{\textbf{Input $m_{3}$}} & \multicolumn{1}{c}{\textbf{Calculated $\rho$}} & \multicolumn{2}{c}{\textbf{NH-QD}} & \multicolumn{2}{c}{\textbf{IH-QD}}\\\cline{3-6}			
				&  & $m_{1}$ & $m_{2}$ & $m_{1}$ & $m_{2}$\\[5pt]\hline
				0.40 & 0.015 & 0.39689 & 0.39699 & 0.40289 & 0.40299\\[5pt]
				0.10 & 0.24 & 0.08674 & 0.8718 & 0.11104 & 0.11136\\[5pt]
				0.08 & 0.375 & 0.06264 & 0.6326 & 0.09340 & 0.09381\\\hline\hline
			\end{tabular}
			\caption{\textit{The absolute neutrino masses in eV, estimated from oscillation data using calculated $\psi=0.031667$ as defined in the text.}}
		\end{center}
	\end{table}
	\newpage
	\section{\label{sec:level3}\Large{Numerical Analysis and Results}}
	For numerical computation of absolute neutrino masses, we take the following observational data: $\triangle m_{12}^{2}=m_{2}^{2}-m_{1}^{2}=7.6\times10^{-5}$ $eV^{2}$, $\left|\triangle m_{32}^{2}\right|=\left|m_{3}^{2}-m_{2}^{2}\right|=2.40\times10^{-3}$ $eV^{2}$; and define the following parameters $\phi=\frac{\left|\triangle m_{23}^{2}\right|}{m_{3}^{2}}$ and $\psi=\frac{\triangle m_{21}^{2}}{\left|\triangle m_{23}^{2}\right|}$, where $m_{3}$ is the input quantity. For NH-QD, the other two mass eigenvalues are estimated from $m_{2}=m_{3}\sqrt{1-\phi}$;  $m_{1}=m_{3}\sqrt{1-\phi(1+\psi)}$ and for IH-QD from  $m_{2}=m_{3}\sqrt{1+\phi}$;  $m_{1}=m_{3}\sqrt{1+\phi(1-\psi)}$.  For suitable input value of $m_{3}$, one can estimate the values of $m_{1}$ and $m_2$ for both NH-QD and IH-QD cases, using the observational values of $\left|\triangle m_{23}^{2}\right|$ and $\triangle m_{21}^{2}$. Table-1 gives the calculated numerical values for two models namely NH-QD and IH-QD for $\left|\triangle m_{23}^{2}\right|=7.6\times10^{-5}$ $eV^{2}$ and $\triangle m_{21}^{2}=2.40\times10^{-3}$ $eV^{2}$.
	\subsection*{\Large{Parameterizations:}} In the next step, we parameterize the mass matrix equation (1) into three types:\\\\
	\textbf{Type IA} with D$=Diag(m_{1},-m_{2},m_{3})$\textbf{:} The mass matrix of this type [13,15] can be parameterized using two parameters ($\epsilon$, $\eta$):
	\begin{equation}
	m_{LL}=
	\begin{bmatrix}
	\epsilon-2\eta & -c\epsilon & -c\epsilon\\
	-c\epsilon & \frac{1}{2}-d\eta & -\frac{1}{2}-\eta\\
	-c\epsilon & -\frac{1}{2}-\eta & \frac{1}{2}-d\eta
	\end{bmatrix} 
	m_{0}
	\end{equation}
	This predicts the solar mixing angle, $\tan\theta_{12}=-\frac{2c\sqrt{2}}{1+(d-1)\frac{\eta}{\epsilon}}$. When $c=d=1.0$, we get the Tri-Bimaximal Mixing(TBM), $\tan 2\theta_{12}=-2\sqrt{2}(\tan^{2}\theta_{12}=0.50)$ and the values of $\epsilon$ and $\eta$ are calculated for both NH-QD and IH-QD cases, by using the values of Table-1 in these two expressions: $m_{1}=(2\epsilon-2\eta)m_{3}$ and $m_{2}=(-\epsilon-2\eta)m_{3}$. We calculate for $\tan^{2}\theta_{12}=0.50$ and $\tan\theta_{12}=0.45$. The solar angle can be further lowered by taking the values $c<1$ and $d>1$ while using the earlier values of $\epsilon$ and $\eta$ extracted using TBM.\\\\
	\textbf{Type-IB} with D$=Diag(m_{1},m_{2},m_{3})$\textbf{:} This type [13,15] of quasi-degenerate mass pattern is given by the mass matrix,
	\begin{equation}
	m_{LL}=
	\begin{bmatrix}
	1-\epsilon-2\eta & c\epsilon & c\epsilon\\
	c\epsilon & 1-d\eta & -\eta\\
	c\epsilon & -\eta & 1-d\eta
	\end{bmatrix}
	m_{3}
	\end{equation}
	This predicts the solar mixing angle,
	\begin{equation}
	\tan 2\theta_{12}=\frac{2c\sqrt{2}}{1+(1-d)\frac{\eta}{\epsilon}}
	\end{equation}
	which gives the TBM solar mixing angle with the input values c=1 and d=1. Like in Type-IA, here $\epsilon$ and $\eta$ values are computed for NH-QD and IH-QD, by using Table-1 in $m_{1}=(1-2\epsilon-2\eta)m_{3}$ and $m_{2}=(1+2-2\eta)m_{3}$. $m_{1}=(1-2\epsilon-2\eta)m_{3}$.\\\\
	\textbf{Type-IC} with D$=Diag(m_{1},m_{2},-m_{3})$\textbf{:} It is not necessary to treat this model [13] separately as it is similar to Type-IB except the interchange of two matrix elements (22) and (23) in the mass matrix in equation (5), and this effectively imparts an additional odd CP-parity on the third mass eigenvalue $m_{3}$ in Type-IC. Such change does not alter the predictions of Type-IB. Our numerical results for both $\tan\theta_{12}=$0.5 and 0.45 cases, in all types of QD models (Type-IA, IB) are consistent with observational bound from cosmological and both NH and IH patterns are valid within quasi-degenerate model. 
	\section{\label{sec:level4}\Large{Conclusion}}
	We have studied the effects of Majorana phases on the predictions of absolute neutrino masses in three types of quasi-degenerate neutrino mass models having both normal and inverted hierarchical patterns within $\mu$-$\tau$ symmetry. These predictions are consistent with data on the mass squared difference derived from various oscillation experiments, and from the upper bound on absolute neutrino masses in $0\nu\beta\beta$ as well as upper bound of cosmology. The results presented in this article will have important implications in the discrimination of neutrino mass models. 
	\section*{\Large{Acknowledgements}}
	This work is carried out through the DST-SERB, Govt. of India, Research project entitled \textbf{\textquotedblleft Neutrino mass ordering, leptonic CP violation and matter-antimatter asymmetry"} vide grant no. \textbf{EMR/2015/001683}.
	
\end{document}